\documentclass[prb,twocolumn,reprint]{revtex4-2}

\usepackage{graphicx}
\usepackage{dcolumn}
\usepackage{bm}
\usepackage{amsmath}
\usepackage{amssymb}
\usepackage[T1]{fontenc}
\usepackage[section]{placeins}
\usepackage{comment}
\usepackage[dvipsnames]{xcolor}
\usepackage{bm,graphicx,hyperref}
\usepackage[dvipsnames]{xcolor}
\usepackage{threeparttable}
\usepackage{chemformula}[2014/04/08]

\newcommand{\be}{\begin{equation}}
\newcommand{\ee}{\end{equation}}

\newcommand{\bpp}{$\beta^{\prime\prime}$}

\begin{document}
\title{Theory of defect-mediated ionic transport in Li$^{+}$, Na$^{+}$ and K$^{+}$ $\beta$ and \bpp\ aluminas}
\author{Suchit Negi}
\affiliation{Centre for Advanced 2D Materials, National University of Singapore, 117546 Singapore}
\affiliation{Institute for Functional Intelligent Materials, National University of Singapore, 117544, Singapore}
\author{Alexandra Carvalho}
\affiliation{Centre for Advanced 2D Materials, National University of Singapore, 117546 Singapore}
\affiliation{Institute for Functional Intelligent Materials, National University of Singapore, 117544, Singapore}
\email{carvalho@nus.edu.sg}
\author{A. H. Castro Neto}
\affiliation{Centre for Advanced 2D Materials, National University of Singapore, 117546 Singapore}
\affiliation{Department of Materials Science and Engineering, National University of Singapore, 117575 Singapore}
\affiliation{Institute for Functional Intelligent Materials, National University of Singapore, 117544, Singapore}


\begin{abstract}
Alkali metal $\beta$/\bpp\ aluminas are among the fastest ionic conductors, yet little is understood about the role of defects in the ion transport mechanism. Here, we use density functional theory (DFT) to investigate the crystal structures of $\beta$ and \bpp\ phases, and vacancy and interstitial defects in these materials.
We find that charge transport is likely to be dominated by alkali metal interstitials in $\beta$-aluminas and by vacancies in \bpp\ aluminas. Lower bounds for the activation energy for diffusion are found by determining the minimum energy paths for defect migration. The resulting migration barriers are lower than the experimental activation energies for conduction in Na $\beta$ and \bpp\ aluminas, suggesting a latent potential for optimization. The lowest activation energy of about 20 meV is predicted for correlated vacancy migration in K \bpp\ alumina.
\end{abstract}    

\maketitle

\setchemformula{kroeger-vink}

\section{Introduction}

All-solid-state batteries are one of the possible solutions to address the current safety and capacity limitations of conventional batteries\cite{kim2015review,li2021advance,janek2023challenges}.
Solid-state systems are able to offer superior chemical and thermal stability compared to Li-ion batteries with traditional liquid electrolytes, which are toxic, flammable and unstable in contact to electrode materials\cite{zhang2017single}. In contrast, some solid-state electrolytes are stable in contact with metallic anodes, allowing for higher energy density storage\cite{dudney2015handbook}. Additionally, they can serve concurrently as ion conduction layers, electronic insulators, and as mechanical separators between the anode and the cathode, allowing for easier design and assembly\cite{kong2021configuring}.

The Na aluminates $\beta$-alumina and $\beta^{\prime\prime}$-alumina combine high conductivity and a wide electrochemical stability window with air processability, making them exceptional materials for Na all-solid-state-batteries\cite{fertig2022high}.
$\beta$-alumina has an ionic conductivity of 0.01-0.03 Scm$^{-1}$ in single crystals at room temperature\cite{lu2010sodium}, of the same order of magnitude as conventional organic liquid electrolytes\cite{valoen2005transport}; $\beta^{\prime\prime}$-alumina solid electrolytes, which are more difficult to grow, can have even higher ionic conductivity despite the lower crystallinity\cite{lu2010sodium,bates1982solid,fertig2022high}. $\beta^{\prime\prime}$-alumina solid electrolytes have been employed in molten-sodium batteries 
as well as planar-type Na-MH batteries\cite{lemmon2016planar}.

The $\beta$/\bpp\-alumina solid electrolytes can also conduct Li, K, Ag and other ions\cite{bates1982solid,BRIANT1980}.
The measured room temperature conductivity for Li is about one order of magnitude lower
than for Na\cite{briant1981ionic}.
It is generally believed that this is due to the fact that while Na$^+$ ions form a plane between the spinel blocks,  Li$^+$ ions hop between positions above and below the conduction plane, leading to slower conduction in the plane.

In contrast, the single-crystal ionic conductivity of K \bpp-alumina has been found to be even higher than that of the Na \bpp-alumina at room temperature by several groups\cite{BRIANT1980,tsurumi1987mixed}. Others, however, found poorer conductivity in ceramic samples, possibly due to the presence of grain boundaries and phase transformations\cite{crosbie1982potassium,williams1992high}.
The increase of the bulk contribution to the impedance was also found to be inconsistent with previous studies\cite{williams1992high}.
Clarifying whether the ionic conductivity of K $\beta^{\prime\prime}$-alumina can achieve such high values as claimed by the earlier studies, and in what conditions, is highly desirable due to possible application of this electrolyte in K-S  batteries\cite{lu2015low}, liquid metal flow batteries\cite{baclig2018high} and other devices\cite{schierle2013solid,Xu2023}.

In this study, we use density functional theory calculations to investigate the basic ion-transport mechanisms in Li, Na, and K $\beta$-aluminas. We will demonstrate that in $\beta$-alumina interstitial mechanisms dominate, whereas in $\beta^{\prime\prime}$-alumina vacancy mechanisms dominate, indicating that control of the occupation of the conduction plane sites is of paramount importance to increase conductivity. We suggest lower bounds for the activation energies for diffusion, indicating that both Na and K $\beta^{\prime\prime}$-aluminas have the potential to offer nearly ideal ion conduction.

\section{Methods\label{methods}}

First-principles DFT calculations were carried out using the {\sc Quantum ESPRESSO} package\cite{giannozzi2017advanced}.
The exchange-correlation functional of Perdew, Burke and Ernzerhof (PBE)\cite{Perdew1996} was used together with ultrasoft pseudopotentials to account for the core electrons\cite{rkkjus}.
We employed a plane-wave basis set with kinetic energy cutoffs of 66~Ry to expand the electronic wave functions. 
The Brillouin zone was sampled using a $\Gamma$-centered 1$\times$1$\times$1 Monkhorst-Pack (MP) grid\cite{mpgrid} for all supercell calculations.

We used the nudged elastic band (NEB)\cite{Jonsson1998,Henkelman2002} method to  find the minimum energy path (MEP) on the potential energy surface (PES). The activation energy for migration was obtained from the difference between the MEP highest saddle point energy and the absolute energy minimum.
NEB calculations were performed between energy minima or between an energy minima and a saddle point derived using symmetry considerations. A total of 9 images were used to construct the MEPs.

\section{Results and Discussion}
\subsection{Crystal Structure\label{sec:structure}}
$\beta$-aluminas present a variety of stoichiometries and are often a mixture of the $\beta$ and $\beta^{\prime\prime}$ alumina phases\cite{nafe2022relationship},
with compositions in the ranges Na$_2$O$\cdot$ $n$Al$_2$O$_3$
(8<$n$<11 for Na $\beta$-alumina) and Na$_2$O$\cdot$ $m$Al$_2$O$_3$ (5<$m$<7 for Na$\beta^{\prime\prime}$-alumina)\cite{fertig2022high}. 
In this section, we describe the structures and compositions used in our models, which typify both $\beta$ and $\beta^{\prime\prime}$ aluminas.

\begin{figure}[ht]
\centering
\includegraphics[width=\columnwidth]{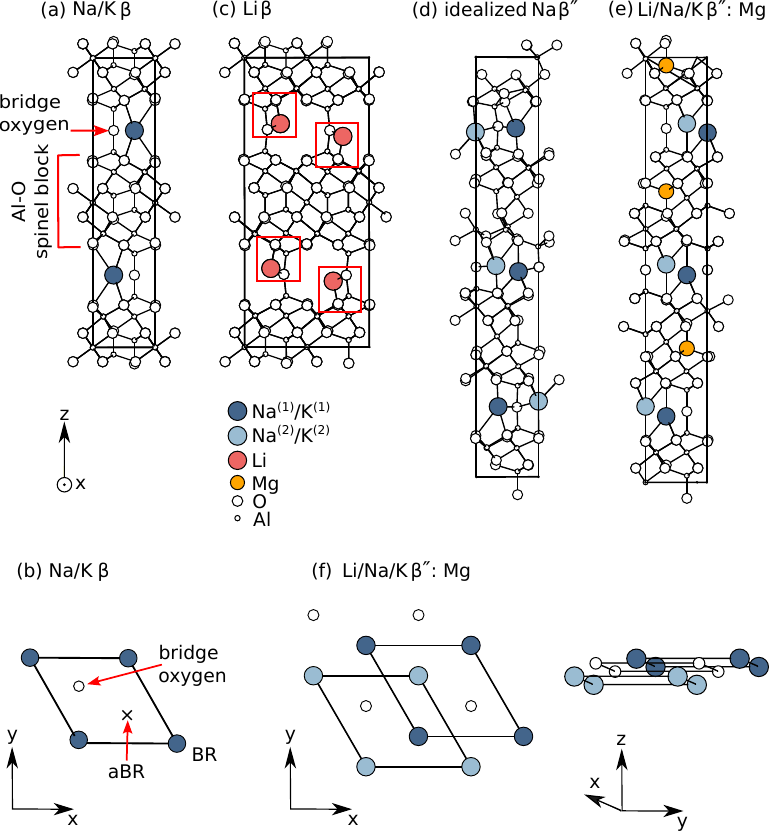}
\caption{Structure of $\beta$- and $\beta^{\prime\prime}$-alumina crystals: (a) unit cell of Na or K $\beta$-aluminas (NaAl$_{11}$O$_{17}$), and (b) detail of the sites in the conduction region; unit cell of (c) Li $\beta$-alumina (LiAl$_{11}$O$_{17}$);(d) idealized $\beta^{\prime\prime}$-alumina (NaAl$_5$O$_8$); (e) Mg-stabilized $\beta^{\prime\prime}$-alumina (Na$_2$MgAl$_{10}$O$_{17}$) and (f) detail of the conduction plane of \bpp\-alumina. The structures of the $\beta^{\prime\prime}$-aluminas are similar for all the three alkali metals. }
 \label{fig:all_aluminas}
\end{figure}

\subsubsection{ Na and K $\beta$ - aluminas}
The nominal phase formula of stoichiometric $X$$\beta$-alumina, where $X$ = \{Li, Na, K\},  is $X$Al$_{11}$O$_{17}$, as determined in the seminal works of Beevers {\it et al.}\cite{beevers1936formula,beevers1937crystal,RAY1975583,bates1982solid}\nolinebreak.
The Na and K $\beta$-aluminas were found to belong to space group 194 ($D^4_{6h}$)\cite{beevers1937crystal}.
Figure~\ref{fig:all_aluminas}-a) shows the Na $\beta$-alumina unit cell containing two formula units (f.u.).
The calculated lattice parameters  can be found in Appendix A~{Tables~\ref{tab:NaKlp},\ref{tab:Lilp}}  and are within 0.2~\AA\ of the experimental values.

The key features of the structure are the Na planes, also referred to as the `conduction region' which alternate with Al-O blocks, also referred to as `spinel blocks'\cite{Boilot1980,RAY1975583,fertig2022high}, by analogy with the MgAl$_2$O$_4$ spinel structure\cite{kummer1972141}. In the spinel block, Al atoms are surrounded by oxygen octahedra or tetrahedra. This block is non-ion conducting and remains nearly undisturbed when the alkali metal ions move. 
Figure~\ref{fig:all_aluminas}-b) shows the Na or K sublattices at the conduction plane. The sites occupied by Na ions are named  `Beevers-Ross' (BR) sites\cite{Boilot1980}.
The unoccupied but crystallographically equivalent site is named `anti Beevers-Ross' (aBR) site\cite{Boilot1980,tsurumi1987mixed}.
All the processes of interest to ion conduction happen in this conduction region.

\subsubsection{Li $\beta$ - alumina}
The Li $\beta$-alumina structure is similar to the Na and K $\beta$-alumina structures [Fig.~\ref{fig:all_aluminas}-c)]. The main difference is that 
The Li atoms are displaced 0.56~\AA~above or below the BR sites, bonding to the neighboring oxygen atoms of the spinel layers immediately above or below, with a consequent doubling of the primitive unit cell along the $\hat{x}$ direction. The corresponding distortion energy is 0.34 eV per unit cell. The resulting space group 18 ($D_2^3$),
as determined with tolerances of 0.1~\AA\ and 0.5$^\circ$ for
distances and angles, respectively. 

 Experimental evidence of the displacement of Li atoms from the BR sites can be found in the frequency of the Raman bands of the Li translational modes, found at 340—410~cm$^{-1}$, higher than expected from the corresponding values reported for other alkali-metal ions\cite{RamanLi}.
 Besides, the probability of occupation of the out-of-plane position by Li can also be derived from an  analysis of NMR \ch{^7Li} quadrupole interactions below 100~K\cite{villa1982lithium,chowdhury2014study}\nolinebreak. An activation energy for out-of-plane motion of 29~meV has been determined in in 93\%Li/7\%Na $\beta$-alumina, consistent with the value 42 meV derived from our calculations\cite{chowdhury2014study}.
 Another study using neutron and X-ray diffraction in single crystals with approximately 61\%Li/39\%Na suggested that the Li atoms are placed 1~\AA~above the BR site at low temperature\cite{edstrom1997li+}. 
 The displacement is larger than predicted by our calculations and could possibly be due to the presence of the larger Na ions in the experiment.
 In contrast, a neutron diffraction study in 50\%Li/50\% sodium $\beta$ alumina\cite{tofield1979structure} determined that the Li atoms sit 1~\AA~ above the mO site at 4.2 K, which is in conflict with our calculations where such position is found to be a saddle point, as will be discussed.

\subsubsection{Idealized \bpp\ aluminas}
The key difference between $\beta$ and $\beta^{\prime\prime}$ aluminas is the stacking sequence of the combined spinel and conduction double layers: in the $\beta$ structure, they have AB stacking, while in the $\beta^{\prime\prime}$ structure they have ABC stacking; the experimental structure belongs to space group 166 ($D^5_{3d}$)\cite{jorgensen1981conduction}.
The idealized phase formula of Na \bpp-alumina is NaAl$_5$O$_8$, of which three f.u. make up a primitive unit cell.
We have not found experimental or theoretical reports of the structure of the ideal stoichiometric $\beta^{\prime\prime}$ phase. 
Rather, $\beta^{\prime\prime}$ alumina is often stabilized by extrinsic divalent cations such as Mg$^{2+}$\cite{fertig2022high}.

 Our idealized structure [Fig.~\ref{fig:all_aluminas}-d)], given as Supplementary Material, is based on the structure of potassium $\beta^{\prime\prime}$-aluminogallate\cite{potassium_cif}, where 30 Al atoms are distributed over 36 Al sites (octahedral and tetrahedral) and 48 O atoms distributed over 51 O sites. For the purpose of the DFT calculations, we have randomly chosen the positions of the Al and O atoms in the spinel block as they have little influence on ion conduction. 
 The unit cell has twice the number of $X$ ions per conduction plane compared to $\beta$ alumina, and instead of forming a planar lattice,  they are staggered 0.22~\AA~or 0.14~\AA~above or below the BR/aBR sites for $X=$Li, Na and K, respectively (Fig.~\ref{fig:all_aluminas}-d).
 
 The Na  $\beta^{\prime\prime}$ in-plane lattice parameter is 1.6\% larger than that of Na $\beta$-alumina, and 1.5\% larger than the experimental value (see Appendix A). The $c$ parameter is 3.2\% smaller than the experimental value\cite{RAY1975583}.
 
\begin{figure}[h!]
\centering
\includegraphics[width=\columnwidth]{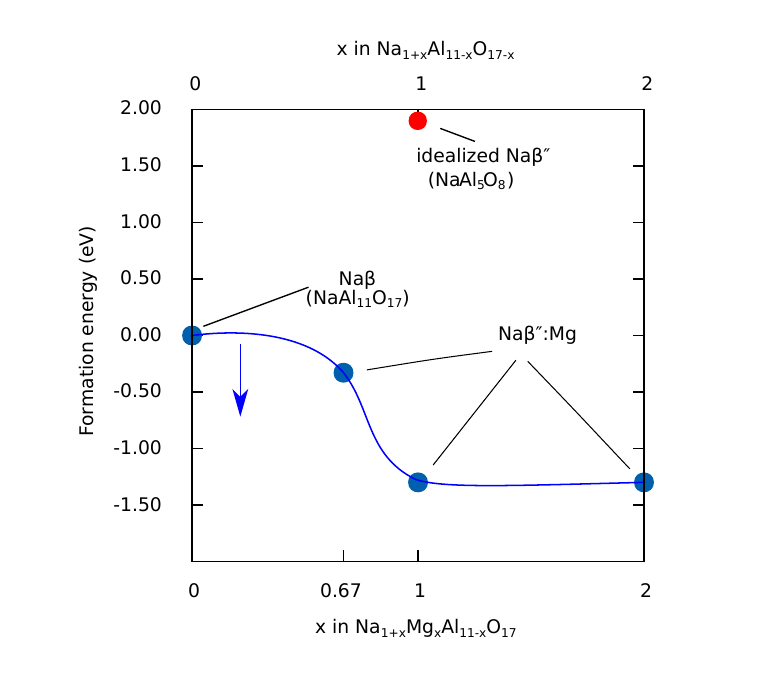}
\caption{Formation energies of idealized Na\bpp\ and Mg-stabilized Na\bpp\ phases relative to Na $\beta$-alumina, in MgO and Na$_{2}$O conditions. 
The blue curve corresponds to the Mg-stabilized Na\bpp\  phase with phase formula Na$_{1+x}$Mg$_{x}$Al$_{11-x}$O$_{17}$. The special case for $x=0$ yields $\beta$ alumina.
The red dot corresponds to the idealized Na\bpp\ alumina phase (NaAl$_5$O$_8$).}
\label{fig:relative_eng}
\end{figure}

 Inspired by the phase diagram of the \ch{Na_2O/Al2O_3} system\cite{sudworth1985sodium}, we compare the formation energy of the idealized Na\bpp\ phase with respect to the Na$\beta$ and $\alpha$ alumina phases, obtaining
 \be \rm
  NaAl_{5}O_{8} (\beta^{\prime\prime}) + 3Al_{2}O_{3} (\alpha) \rightarrow NaAl_{11}O_{17} (\beta) + 1.9\:eV.
 \ee
 Thus the undoped Na\bpp\ phase is unstable (Fig.~\ref{fig:relative_eng}).
Moreover in the calculated DFT bandstructure  of idealized Na\bpp\ is $p$-type doped, with the Fermi level at the valence band top, possibly due to the Al vacancies (see Appendix~\ref{appendixa0Eg}).
This justifies theoretically the need to dope the material with stabilizing species. These have to be taken into account in our model to reproduce the electron insulating behavior and the right defect charge states, and will be considered in the next sub-section.

\subsubsection{Mg-stabilized \bpp\ aluminas}

 Divalent cations such as Mg$^{2+}$ act as stabilizers of the \bpp\ phase by compensating the charge of the additional $X$ ions without the necessity to change the spinel layer structure\cite{fertig2022high,Tsurumi,Frase}.
The ideal formula for Mg-stabilized Na \bpp-alumina is 
Na$_{1+x}$Mg$_{x}$Al$_{11-x}$O$_{17}$ [Fig.~\ref{fig:all_aluminas}e-f)] where the Mg content $x$ can be varied while keeping the number of additional Na atoms equal to the number of substitutional Mg at Al sites (Mg$_{\rm Al}$) so as to keep charge neutrality.
We estimate $x=2$ with three Na atoms per unit cell and per plane to be the highest possible Na packing, with a minimum Na-Na distance of 2.9~\AA, comparable to the double of the ionic radius of Na$^+$ (1.02~\AA). Experimentally, $x\sim 0.66$ has been reported\cite{collin1986host,Boilot1980}, and possibly an imbalance of Na compared to Mg due to sodium evaporation above 1600$^\circ$C\cite{lu2010sodium}.

Adding a small fraction of Mg lowers the formation energy relative to Na$\beta$ phase (Fig.~\ref{fig:relative_eng}), thus making the Mg-stabilized Na\bpp\ phase more favorable over the Na$\beta$ phase. While the formation energy of idealized Na\bpp\ is positive, that of the Mg-stabilized Na\bpp\ phases is now negative. Mg-stabilized Na \bpp-aluminas where $x$=1 and $x$=2 in the formula Na$_{1+x}$Mg$_{x}$Al$_{11-x}$O$_{17}$ are insulators with a clean bandgap (see Appendix~\ref{appendixa0Eg} Fig~\ref{fig:bands}).

Figure~\ref{fig:all_aluminas}-e) shows the structure for $x=1$, with composition Na$_2$O$\cdot$MgO$\cdot$$5$Al$_2$O$_3$, of which three f.u. make up a primitive unit cell. The three Mg$_{\rm Al}$ are placed one in each spinel block and as far as possible from the conduction planes. 
 The resulting structure resembles that of undoped Na \bpp-alumina (Fig.~\ref{fig:all_aluminas}-c,d), with the same stacking sequence, and similar up and down staggering of the Na ions in the conduction region, which occupy all the equivalent BR/aBR sites.
  Its optimized lattice parameters are given in Appendix A Table~\ref{tab:NaKlp}.
  The structures of Mg-doped Li and K \bpp-aluminas are very similar to that of Na \bpp-alumina.
  The structure with $x=1$ has the same number of $X$ ions in every conduction plane and therefore we will use it as a model in subsequent calculations for all \bpp-aluminas, unless otherwise stated.

\subsection{Defect Structures}

\subsubsection{Vacancy}

We created a single vacancy in one of the conduction planes, for each of the materials studied studied. 

In the Li $\beta$-alumina, the removal of the Li$^+$ ion leads to an expansion of the distance between its two nearest bridge oxygen neighbors in the conduction region by 18\%, whereas the neighboring Li remains bonded to the respective oxygen neighbors.
In the Na and K $\beta$ aluminas, the vacancies retain the trigonal symmetry of the original sites, with the triangle of bridge oxygen nearest neighbors, in the same (0001) plane, contracting by 18\% and 8\%, respectively.

\begin{figure}[h!]
\centering
\includegraphics[width=\columnwidth]{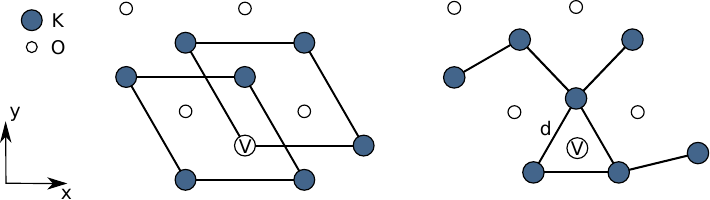}
\caption{Schematic representation of a vacancy in Mg-stabilized Na or K \bpp\ alumina before relaxation (left) and after relaxation (right).
`V' indicates a vacant site. Only atoms at the conduction plane are shown for clarity.}
\label{fig:K_prime_vacancy}
\end{figure}

In the \bpp\ aluminas, the distance between the V$_X'$ nearest $X$ neighbors ($d$) contracts by 2\%, 54\% and 30\% for $X$=Li, Na and K, respectively.
The structure of the reconstructed vacancy is shown in Fig.~\ref{fig:K_prime_vacancy}. In the resulting structure, three $X$ atom rings form around the vacant site, and the adjacent rings of $X$ atoms around the oxygen sites become five-atom rings instead of six-atom rings. This planar arrangement is more prominent in K with effective bond length $d\sim$ 4~\AA~ compared to K$^+$ ionic radii $\sim$ 1.38~\AA. 

In the case of the $\beta$ phases, the presence of the vacancy disturbs the positions of the Na atoms in the other conduction plane as well. In the case of the \bpp\ phases however, the relaxation in the other conduction planes was negligible. In all cases changes to the spinel structures are negligible.

\begin{figure}[h!]
\centering
\includegraphics[width=\columnwidth]{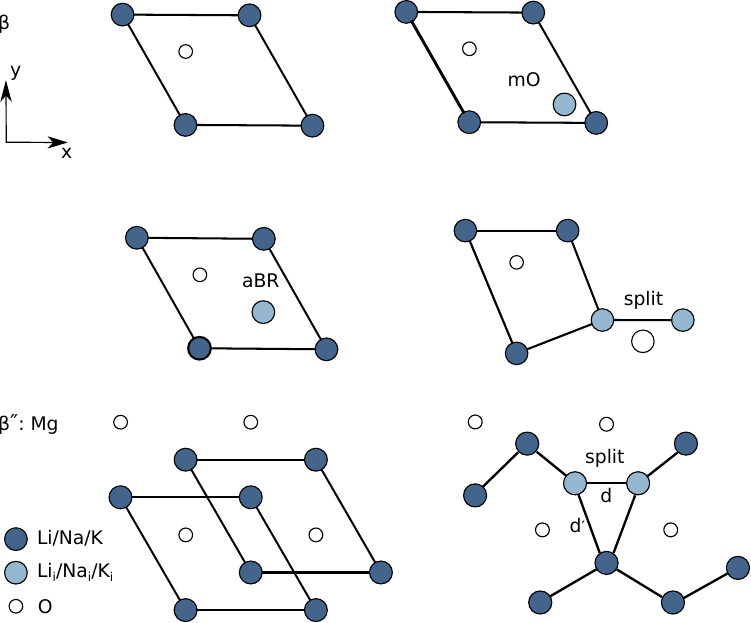}
\caption{Schematic representation of the interstitial sites at the conduction planes
of $\beta$ alumina (top) or \bpp\ alumina (bottom).
The mO and aBR sites are unstable with respect to relaxation to the split-interstitial configuration. The split-interstitial configurations are shown at the relaxed geometries obtained for Na $\beta$/\bpp\ aluminas. 
Only atoms at the conduction plane are shown.}
 \label{fig:Na_beta_interstitial}
\end{figure}

\subsubsection{Interstitial}
As considered in sec.~\ref{sec:structure}, the conduction planes of both $\beta$ and \bpp\ aluminas have
four sites equidistant to the bridging oxygens.
In $\beta$ alumina, two of these are occupied by $X$ (the BR sites), and two are unoccupied (the aBR sites) -- see Fig.~\ref{fig:Na_beta_interstitial}. However, in \bpp\ alumina all four sites  are occupied by $X$ and are crystallographically equivalent.
Additionally, there is another high-symmetry interstitial site named the mid-oxygen (mO) site, equidistant from an aBR site and a BR site. Lastly, we have considered split-interstitials consisting of two $X$ atoms at adjacent mO sites, replacing the original atom at the BR site. 

In Li $\beta$-alumina, the lowest energy configuration is a distorted [1$\bar{1}$0] split-interstitial, with two Li atoms approximally situated between the original Li site and the aBR site.
In Na  $\beta$-alumina, a $\langle 10\bar{1}0\rangle$ split-interstitial is the only stable configuration.
Both the aBR interstitial and the mO interstitial relax to the split interstitial configuration. In K $\beta$-alumina, the $\langle 10\bar{1}0\rangle$ split-interstitial and the aBR interstitial are distinct but  degenerate in energy.

Experimental studies on Na $\beta$ alumina suggest Na ions start occupying the mO site in Na-rich $\beta$-alumina\cite{WOLF1979757,Roth1976,collin1986host}.
This is consistent with the results of our calculations, since in the split-interstitial configuration, both Na atoms are only 0.4~\AA\ away from the respective nearest mO sites.

As for the Mg-stabilized Na \bpp-alumina, Na ions occupy all the available BR/aBR sites in the conduction region. Thus only the vicinity of mO interstitial sites is available for interstitial Na ions, as shown in Fig.~\ref{fig:Na_beta_interstitial}. The split-interstitial, with two atoms occupying adjacent mO sites instead of the original BR/aBR site, was found to be the only stable structure.
The presence of the Na interstitials has little influence on the atoms outside the conduction region.

\subsection{Defect Formation Energies}

We now investigate the role of vacancy and interstitial defects in the ionic diffusion and charge conduction. If the defects are thermally generated, in equilibrium conditions, the activation energy for conduction is the sum of two terms -- the energy for defect formation plus the energy for defect migration. 

However, defects may be present due to non-adiabatic processes during growth.
For example, the \bpp-aluminas are usually Na-deficient, with the presence of Na vacancies\cite{engstrom1981ionic}, and this is believed to result from Na evaporation above 1600$^\circ$C\cite{lu2010sodium}.
In such material, the activation energy for conduction measured in a closed system is the migration energy only. Thus, migration energies are the minimum bound for the activation energy. 

In an alkali metal battery context, the alkali metal chemical potential can vary across the electrolyte due to the proximity to the cathode or anode. Here, we calculate the formation energy of alkali metal vacancies (V$_X$) and interstitials ($X_{\rm i}$) in $X$-rich conditions, where $X$ = \{Li, Na, K \}. 

\begin{figure}[h!]
\centering
\includegraphics[width=\columnwidth]{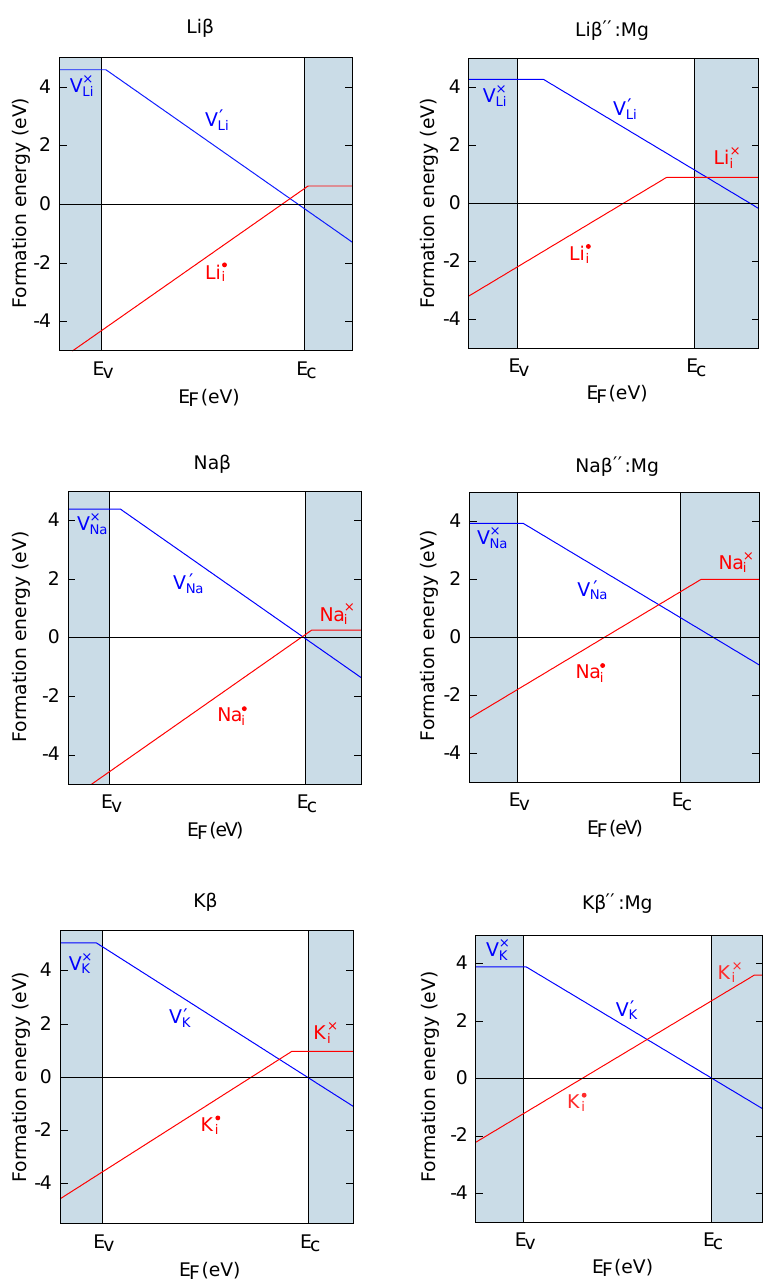}
\caption{Vacancy and interstitial formation energies in $X$-rich conditions ($X$ = \{Li, Na, K \}). 
The formation energies for defects in the $\beta$-alumina phases are shown on the left, and those in the Mg-stabilized \bpp\ alumina phases are shown on the right.
The bandgap values were calculated and can be found in Appendix (Tables~\ref{tab:NaKlp},\ref{tab:Lilp}).
}
\label{fig:formation_eng}
\end{figure}

Even though $\beta$-aluminas are wide gap insulators, we assume that its Fermi level ($E_F$) is well defined and take it to be the chemical potential for electrons. The defect formation energy in relative charge state $q$ is then given by
\be E_{\rm f}(D^q) = E_{\rm t}(X\beta:D^q) - E_{\rm t}(X\beta) \pm E_{\rm t}({\rm bcc}\mbox{-}X)+qE_F,\ee
where $E_{\rm t}(X\beta:D^q)$, $E_{\rm t}(X\beta)$ and $E_{\rm t}({\rm bcc}\mbox{-}X)$ are the total energies of the supercell with the defect, the pristine supercell and the metallic reservoir of element $X$, respectively, and the $-/+$ signs are for interstitial/vacancy defects.
The results for vacancies and interstitials are shown in 
Fig.~\ref{fig:formation_eng}.

The vacancy $(-/0)$ transition level is close to the top of the valence band ($E_v$), indicating that vacancies are always negatively charged (\ch{V_{\it X}'}). Similarly the $(0/+)$ transition levels of interstitial defects are close to the conduction band ($E_c$) indicating that interstitials are always positively charged (\ch{{\it X}_i^*}). 
Thus both $X$ vacancies and interstitials can in principle be responsible for ionic charge conduction. 

In the case of the $\beta$-aluminas, if the Fermi level is close to the conduction band, both vacancies and interstitials can be created with nearly vanishing or negative formation energy, but interstitials are more favorable over a wide range of Fermi level energies.

In the case of the \bpp\ aluminas, the formation energies of interstitial defects are higher than in the corresponding $\beta$-aluminas, because the conduction plane is more densely packed.
However, neutral vacancy formation energies are still high, around 4~eV in $X$-rich conditions.

The definition of the chemical potentials in $X$-poor conditions in battery systems depends on the electrodes used.
In NaS batteries, Na-poor conditions can be defined by assuming equilibrium with a reservoir of the sodium polysulfide \ch{Na_2S_4} and sulfur, which results in vacancy formation energies that are about 2~eV lower than in Na-rich conditions. 
In general, battery voltages are of the order of $\sim$ eV, corresponding to the difference in the $X$ chemical potential at the $X$-poor side and at the $X$-rich side of the battery.

Finally, we note that the calculated bandgap is underestimated as expected using the DFT-PBE functional. Assuming that the defect levels are pinned to the band edges, correcting the bandgap could lead to vanishing defect formation energies near mid-gap.

\begin{figure}[h!]
\centering
\includegraphics[width=\columnwidth]{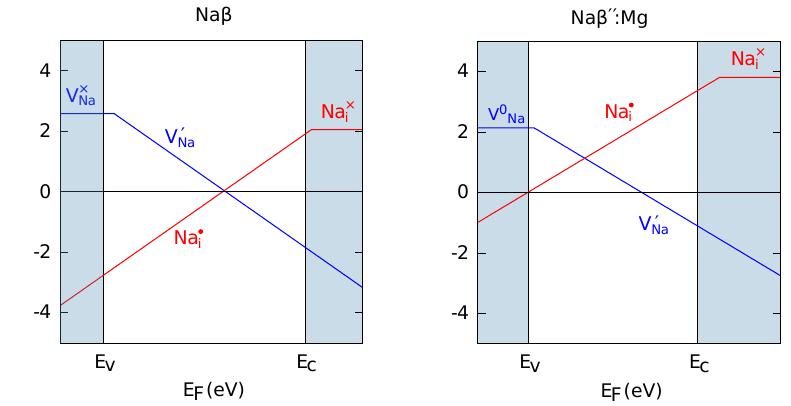}
\caption{Vacancy and interstitial formation energies in Na-poor conditions (with chemical potentials defined by \ch{Na_2S_4} and \ch{S_8} reservoirs). 
The formation energies for defects in the $\beta$-alumina phases are shown on the left, and those in the Mg-stabilized \bpp\ alumina phases are shown on the right.
The bandgap values were calculated and can be found in Appendix (Tables~\ref{tab:NaKlp},\ref{tab:Lilp}).}
 \label{fig:Na_poor}
\end{figure}

\newpage
\subsection{Migration Energies}
The activation energies for migration of $X$ vacancies and interstitials have been calculated assuming that they are always in their respective charged states (\ch{V_{\it X}'} and \ch{{\it X}_i^*}).
The activation energies are found by searching the minimum energy path between two equivalent positions separated by a lattice vector, using the NEB method (see Methods section), and taking the difference between the energies at the saddle point and at the minimum energy point.

\subsubsection{Vacancy Migration}
\begin{figure}[h!]
\centering
\includegraphics[width=0.97\columnwidth]{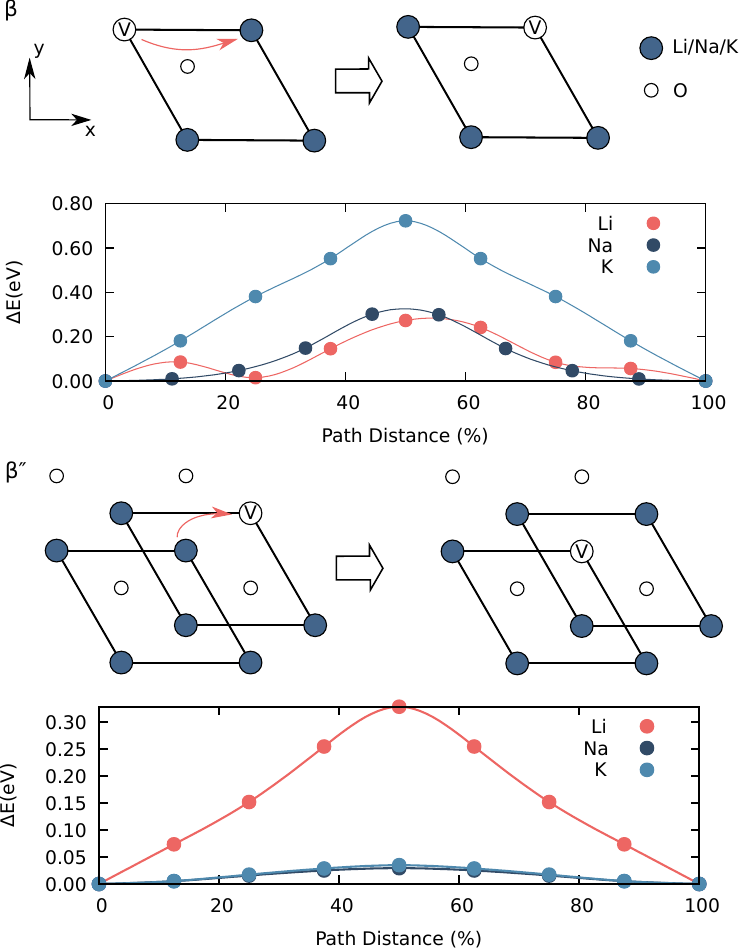}
\caption{Schematic of the vacancy migration pathways in $\beta$-aluminas (top) and Mg-stabilized \bpp-aluminas (bottom) along with their respective energy profiles. }
\label{fig:vacancy_migration}
\end{figure}

\paragraph{$\beta$-alumina}
Figure~\ref{fig:vacancy_migration} shows a hop of the Na vacancy in Na $\beta$-alumina from one BR site to another, and the corresponding energy profile.
The saddle point is at the middle plane equidistant to the initial and final configurations.
The migration path is similar for K $\beta$-alumina.
In Li $\beta$ alumina, the Li vacancy is not trigonal, due to the original position of the Li atoms closer to the bridge oxygen atoms. The Li vacancy migration involves a jump of one of its Li neighbors to the bridge oxygen atom near the vacancy, breaking two Li-O bonds, but still requires a lower activation energy than in the cases of Na or K.
The respective activation energies can be found in Table~\ref{tab:defect_migration}. 

\paragraph{Mg-stabilized \bpp\ aluminas}
In the \bpp\ conduction region, the $X$ atoms form a honeycomb lattice and are $1/\sqrt{3}a_0$ apart, closer than in $\beta$-alumina. The migration energy for the vacancy hop is 0.33, 0.03 and 0.03 eV for Li, Na and K, respectively (Table~\ref{tab:defect_migration}).
The small migration energies obtained for Na and K are already close to the uncertainty of our calculations.

We have also investigated the two-atom correlated process where one atom jumps into the neighboring vacancy and its neighbor jumps to the site just vacated (Fig.~\ref{fig:vacancy_migration}) has an activation energy of just 0.32, 0.08 and 0.02 for Li, Na and K respectively. The concerted movement allows two atoms to move with approximately the same energy as one atom for Li and K. It is possible that concerted migration involving more atoms is also energetically favored, but we could not investigate it due to the size limit imposed by the supercell dimensions.

Since the \bpp\ aluminas are often alkali metal deficient, these migration energies can be considered lower bounds for the activation energy for the conductivity, which have been experimentally determined to be 0.30, \textcolor{blue}{0.03} (at high temperature) and 0.15 eV in Li, Na and K \bpp\ aluminas, respectively\cite{bates1982solid}.

The exceptionally low activation energy found for \ch{V_K'} in K \bpp: Mg corroborates the experimental observation of room temperature ionic conductivity of K \bpp\ phase\cite{BRIANT1980} being 2000 times higher than K $\beta$ and 10 times higher than that of Na \bpp-alumina. 

\subsubsection{Interstitial Migration}

Since the $X$ interstitials in all $\beta$ and \bpp\ alumina materials considered are split-interstitials, their migration necessarily involves at least two atom jumps.

\begin{figure}[h!]
\centering
\includegraphics[width=\columnwidth]{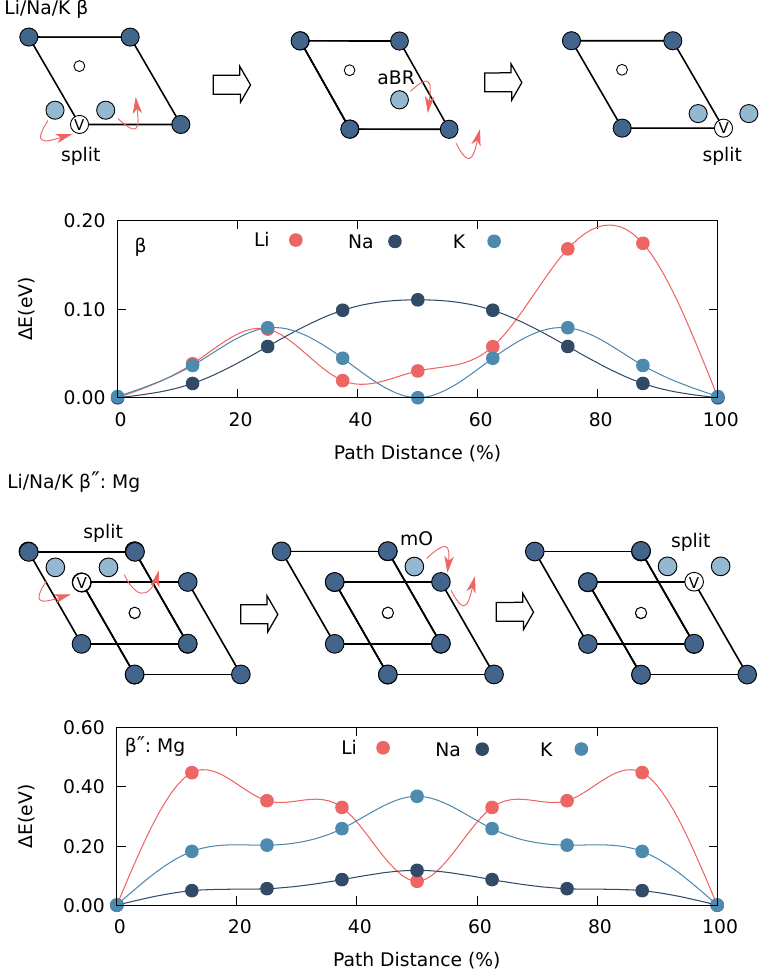}
\caption{Schematic representation of Li, Na and K interstitial migration in the respective $\beta$-alumina (top) and Mg-stabilized \bpp-alumina (bottom), alongside the calculated energy profile.
Different from all other $X$ self-interstitials which have the orientation represented in the figure, the Li split-interstitial in Li $\beta$-alumina is oriented approximately along the [110], but follows a similar migration path.
The migration path distance is from a split-interstitial position to a neighboring split-interstitial position (at 0 and 100\%), through the aBR site in the case of $\beta$-alumina, or through the mO site in the case of Mg-stabilized \bpp\ alumina.
For orientation and color code, please refer to Fig.~\ref{fig:Na_beta_interstitial}.
}
 \label{fig:interstitial_migration}
\end{figure}

\paragraph{$\beta$-aluminas}
Figure~\ref{fig:interstitial_migration} shows schematically the migration of a split-interstitial in $\beta$-alumina along one of the principal in-plane crystal directions. The saddle point is an interstitial at the  aBR site, a configuration at the mirror plane equidistant to the initial and final configurations. While one of the split-interstitial atoms hops to the vacant site, the other one hops through the aBR site, knocking on the next $X$ atom to create another split-interstitial one lattice spacing away.

The energy profile shows a monotonic increase from the split-interstitial to the aBR site for Na but not for Li and K (Fig.~\ref{fig:interstitial_migration}).
The migration energies can be found in Table~\ref{tab:defect_migration}. 
The interstitial migration energies are lower than the vacancy migration energies for all $\beta$-alumina. 
Coupled to the low $X$ interstitial formation energy
this indicates that an interstitial-mediated mechanism is the dominant process of ionic conduction.
The activation energies 0.17, 0.09 and 0.08 eV for Li, Na and K, respectively, can be considered lower bounds for the activation energy for conduction, which has been experimentally measured to be 0.24, 0.15 and 0.28~eV, respectively\cite{bates1982solid}.

\paragraph{Mg-stabilized \bpp\ aluminas}
As for Mg-stabilized \bpp\ aluminas, 
the $X$ interstitials migrate through  the available mO sites (Fig.~\ref{fig:interstitial_migration}).
One of the atoms in the split interstitial migrates to the vacant site while repelling the other, which knocks-on one of its nearest neighbors into an available mO site. This mO interstitial then can relax to a split-interstitial centered on another lattice site.
 The process is not always on the plane, as the $X$ atoms in the two BR/aBR sublattices are staggered, and more packed than in the $\beta$-aluminas.
The respective activation energies for this process are 0.45, 0.12 and 0.37~eV for Li, Na and K, respectively. All these are higher than the respective vacancy migration energies (Table~\ref{tab:defect_migration}).
The energy profiles for the interstitial migration in the three \bpp\ materials can be found in Fig.~\ref{fig:interstitial_migration}. 

\begin{table}[htb]
\caption{Calculated activation energies for migration ($W_{\rm mig}$) of alkali metal vacancy and interstitial defects, and experimental activation energy derived from conductivity experiments ($E_{\rm a}$). Values in square brackets are for a correlated two-atom migration. LT and HT refer to low temperature and high temperature, respectively. Values in itallic were reported for single crystals. The calculated and experimental values can be directly compared if the material has pre-existing defects of either type, or in thermodynamic equilibrium if the formation energy of one of the defects is zero or negative.\label{tab:defect_migration}}

\begin{ruledtabular}
\begin{tabular}{llll}
Host& \multicolumn{2}{c}{calc. $W_{\rm mig}$} & exp. $E_{\rm a}$ \\
&  V$_X^\prime$  &  $X$\ch{_{i}^{*}} &\\
\hline
Li$\beta$ & 0.27   &0.17 & {\it 0.24}$^a$ \\
        &          &     &{\it 0.27}$^{e}$\\
Li\bpp: Mg & 0.33  &0.45 & 0.30$^b$\\
           & [0.32]\\

Na$\beta$ & 0.30   &0.09 & 0.15$^b$\\
           &       &     & {\it 0.16}$^a$ \\
Na\bpp: Mg & 0.03  &0.12 & 0.28-0.33$^c$ (LT) \\
           & [0.08]   &     & {\it 0.20-0.31}$^g$ (LT) \\
           &       &     & 0.03$^b$ (HT)\\
           &      &     & 0.09-0.12$^g$ (HT)\\
K$\beta$  & 0.72   &0.08 & 0.28-0.56$^d$\\
K\bpp: Mg & 0.03   &0.37 & 0.15$^b$\\
          &  [0.02]&     &{\it 0.186}$^f$\\
\end{tabular}
\end{ruledtabular}
\begin{flushleft}
$^a$~Ref.~\onlinecite{briant1981ionic}\\
$^b$~Ref.~\onlinecite{bates1982solid}\\
$^c$~Ref.~\onlinecite{engstrom1981ionic}\\
$^d$~Ref.~\onlinecite{allen1978far}\\
$^e$~Ref.~\onlinecite{chowdhury2014lithium,chowdhury2014study}\\
$^f$~Ref.~\onlinecite{williams1992high}\\
$^g$~Ref.~\onlinecite{BRIANT1980}
\end{flushleft}
\end{table}

\section{Conclusion}

We have carried out DFT calculations of the structure, formation energy and migration of intrinsic defects in $\beta$ and \bpp\ aluminas of Li, Na and K.
We have confirmed that both alkali metal self-interstitials and vacancies in the conduction plane can carry charge almost for the whole range of Fermi level energies potentially available.

Alkali metal self-interstitials have low or even negative formation energy in $\beta$-aluminas
in alkali metal-rich conditions, and this is consistent with Na excess reported experimentally\cite{allen1978far}.
The $X_i$ activation energies for migration are slightly lower than the experimental activation energies, which could be due to the formation energy contribution or to crystal imperfection.

The formation energies of alkali metal vacancies in \bpp-aluminas are positive over most of the DFT bandgap even in the case of Na-poor Na \bpp-alumina. However, such defects can possibly be present due to high-temperature processing. Similarly, the structure can be made less Li/Na/K stuffed (but stoichiometric) by adding less \ch{Mg_{Al}}.

In the $\beta$-aluminas, alkali metal vacancy migration energies increase with increasing ionic radius, but the alkali metal interstitial migration energy decreases with ionic radius, which is somewhat counterintuitive.  In the \bpp-aluminas, the vacancy migration energies for Na and K are about one order of magnitude lower than for Li. The atomic radii of Na and K are just ideal to move in a snug fit in the interlayer spacing when it is close to maximum packing. 

The vacancy migration energy that we have obtained for Na \bpp-alumina is closer to the experimental value for the high-temperature regime.
It has been suggested that the higher activation energy at low temperatures is due to vacancy ordering\cite{engstrom1981ionic,bates1981composition}. However, in our calculations, the vacancies are reconstructed and due to the periodic boundary conditions have a 2$\times$2 ordering without jeopardizing the low activation energy. Recently, it has been proposed that interaction between the vacancies and the \ch{Mg_{Al}} is at the origin of higher activation energy in the low-temperature regime\cite{poletayev2022defect}, an explanation that reconciles the experimental interpretation with the results of our calculations.
Comparing the energy of the correlated two-atom process with the one-atom process for vacancy migration in \bpp-aluminas, we found that the energy of the two is nearly the same in the cases of Li and K, indicating that the uphill movement of one atom on its potential energy surface is correlated with the downhill movement of the other atom\cite{he2017origin}. Such processes involving two or more atoms may contribute significantly to diffusion. Unfortunately, there is lack of measurements of the Haven ratio in \bpp-aluminas.

The calculated activation energy for \ch{V_K'} in K \bpp-alumina is found to be only about 20 meV, suggesting that this material can present nearly ideal ion conduction if the amount of K is carefully controlled.
The smaller energy barrier found in K \bpp-alumina is due to its optimal ionic radius together with lower affinity of K for O when comparing to the other alkali metals (see Table~\ref{tab:reaction_eng}). We therefore believe that K \bpp-alumina deserves further attention as an ionic conductor.

\acknowledgements
This research project is supported by the Ministry of Education, 
Singapore, under its Research Centre of Excellence award to the Institute for Functional Intelligent Materials, National University of Singapore
(I-FIM, project No. EDUNC-33-18-279-V12). 
This work used computational resources of the supercomputer Fugaku provided by RIKEN (Project ID: hp230186);  the Centre of Advanced 2D Materials (CA2DM), funded by the 
National Research Foundation, Prime Ministers Office, Singapore; and the Singapore National Supercomputing Centre (NSCC).

\appendix

\section{Calculated lattice parameters and bandstructures \label{appendixa0Eg}}

The calculated lattice parameters $a$, $b$ and $c$ are given in Tables~\ref{tab:NaKlp} and \ref{tab:Lilp}. For hexagonal systems, $\bf a$ and $\bf c$ are aligned with $\hat{x}$ and $\hat{z}$ directions respectively. The respective crystallographic information files (CIF) are given as Supplementary Information.

For direct comparison, we represent the orthorhombic structure of the Li$\beta$ phase in the same orientation, with $\bf a$  and $\bf c$ lattice vectors aligned with the $\hat{x}$ and $\hat{z}$ directions respectively. 
The {\bf b} vector is perpendicular to {\bf c} and makes an angle of 120$^\circ$ with {\bf a}. For a standard crystallographic representation, refer to the CIF file in Supplementary Information.

The electronic bandstructures and bandgaps were obtained using DFT in the PBE approximation as detailed in Section~\ref{methods}.

\begin{table}[htb]
\caption{Lattice parameters and electronic bandgaps of Na and K $\beta$-aluminas, idealized \bpp-aluminas and Mg-stabilized \bpp-aluminas ($x=1$). Experimental values are given in brackets.\label{tab:NaKlp}}
\begin{ruledtabular}
\begin{tabular*}{8.5cm}{llll}
 Structure &  $a$ (\AA)  &  $c$ (\AA)  &   bandgap (eV) \\
\hline
Na$\beta$ & 5.597 (5.594$^{\rm a}$)& 22.485 (22.53$^{\rm a}$)  & 4.61 \\
Na\bpp & 5.680 (5.60$^{\rm a}$)& 34.059 (34.11$^{\rm a}$) &  1.98\\ 
Na\bpp: Mg & 5.698 & 33.784 & 3.58\\
\hline
K$\beta$ & 5.597 (5.61$^{\rm b}$) & 22.527 (22.75$^{\rm b}$)&  4.80\\
K\bpp\ & 5.689 (5.595$^{\rm a}$) & 34.689 (34.226$^{\rm a}$) & 1.90\\
K\bpp: Mg & 5.704 & 34.421& 4.21\\
\end{tabular*}
\end{ruledtabular}
\begin{tablenotes}\footnotesize
\item{$^{\rm a}$ Ref.~\cite{RAY1975583}}
\item{$^{\rm b}$ Ref.~\cite{takahashi1981synthesis}}
\end{tablenotes}
\end{table}

\begin{table}[htb]
\caption{Lattice parameters and electronic bandgaps of Li $\beta$-alumina, idealized \bpp-alumina and Mg-stabilized \bpp-alumina ($x=1$).\label{tab:Lilp}}
\begin{ruledtabular}
\begin{tabular*}{8.4cm}{lccccccc}
 Structure &  $a$ (\AA)  &  $b$ (\AA)&  $c$ (\AA) &  bandgap (eV)\\
\hline
Li$\beta$  & 11.192&  5.596 & 22.462  & 4.63\\
Li\bpp\    & 5.683 & -      & 34.354  & 2.09\\
Li\bpp: Mg & 5.670 & -      & 34.544 & 3.68\\
\end{tabular*}
\end{ruledtabular}
\end{table}

\begin{figure}[ht]
\includegraphics[width=\columnwidth]{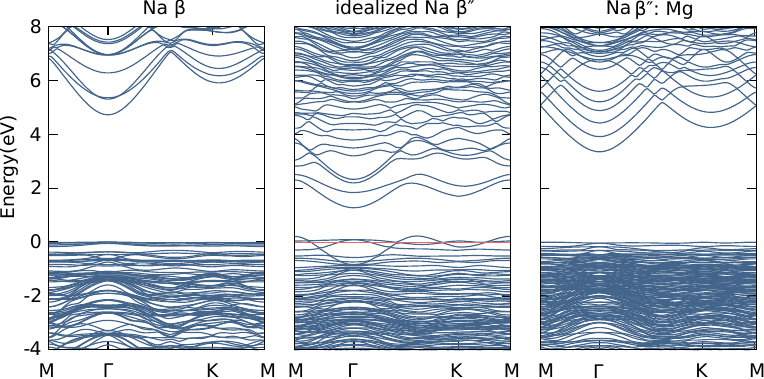}
\caption{Electronic bandstructures of Na$\beta$, idealized Na\bpp\ and Mg-stabilized Na\bpp ($x=1$) phases. For Na $\beta$-alumina and Mg-stabilized Na \bpp-alumina, the valence band maxima was set to zero. For the idealized Na \bpp-alumina, the Fermi level is set to zero.}
 \label{fig:bands}
\end{figure}

\section{Cohesive energies of alkali metal oxides}
Cohesive (atomization) energies of the ground state oxides are given in Table~\ref{tab:reaction_eng}.
\begin{table}[htb]
\caption{Cohesive energies (eV) of lithium, sodium and potassium oxides.\label{tab:reaction_eng}}
\begin{ruledtabular}
\begin{tabular*}{4.5cm}{lc}
Compound & Cohesive energy (eV) \\
\hline
Li$_{2}$O & $-$5.93\\
Na$_{2}$O &  $-$4.02\\
K$_{2}$O  &  $-$3.47\\
\end{tabular*}
\end{ruledtabular}
\end{table}

\bibliography{refs.bib}
\end{document}